\documentclass[runningheads]{llncs}
\usepackage[T1]{fontenc}
\usepackage{graphicx}
\usepackage{cite}
\usepackage{amsmath,amssymb,amsfonts}
\usepackage{algorithmic}
\usepackage{textcomp}
\usepackage{xcolor}
\usepackage{url}
\usepackage{siunitx}
\usepackage{wrapfig}
\usepackage{worldflags}
\usepackage{fontawesome5}
\usepackage{floatrow}

\begin{document}

\title{The Dawn of Decentralized Social Media: An Exploration of Bluesky's Public Opening}

\titlerunning{An Exploration of Bluesky's Public Opening}

\author{Erfan Samieyan Sahneh\inst{1}\orcidID{0009-0004-4924-2315} \and
Gianluca Nogara\inst{1}\orcidID{0000-0002-4412-131X} \and  Matthew R. DeVerna\inst{2}\orcidID{0000-0003-3578-8339} \and Nick Liu\inst{2} \and Luca Luceri\inst{3}\orcidID{0000-0001-5267-7484} \and
Filippo Menczer \inst{2}\orcidID{0000-0003-4384-2876} \and Francesco Pierri\inst{4}\orcidID{0000-0002-9339-7566} \and
Silvia Giordano\inst{1}\orcidID{0000-0003-2603-9029}}
\authorrunning{E. Samieyan Sahneh et al.}
\institute{ISIN - DTI, SUPSI, Lugano, Switzerland \email{erfan.samieyan@supsi.ch} \and
Observatory on Social Media, Indiana University,
 Bloomington, USA
 \and
Information Sciences Institute, USC,
 Los Angeles, California, USA
 \and Dip. Elettronica, Informazione e Bioingegneria, Politecnico di Milano, Italy
 }

\maketitle              

\begin{abstract}
Bluesky is a Twitter-like decentralized social media platform that has recently grown in popularity. After an invite-only period, it opened to the public worldwide on February 6th, 2024. In this paper, we provide a longitudinal analysis of user activity in the two months around the opening, studying changes in the general characteristics of the platform due to the rapid growth of the user base.
We observe a broad distribution of activity similar to more established platforms, but a higher volume of original than reshared content, and very low toxicity. After opening to the public, Bluesky experienced a large surge in new users and activity, especially posting English and Japanese content. In particular, several accounts entered the discussion with suspicious behavior, like following many accounts and sharing content from low-credibility news outlets. Some of these have already been classified as spam or suspended, suggesting effective moderation. 
\end{abstract}

\keywords{Bluesky \and decentralization \and online social media \and misinformation.}

\section{Introduction}

Bluesky Social\footnote{\url{https://bsky.social/about}} is a novel decentralized social media platform for microblogging based in the United States. Originally based on invite-only subscription, Bluesky officially opened to the public on February 6th, 2024 \cite{Silberling:Bluesky-opening}. With only a few thousand users active in January 2024, the platform registered more than one million new users on the first day of its opening \cite{Bluesky2024}.

Bluesky is built upon the Authenticated Transfer (AT) protocol \cite{Kleppmann2024ATProtocol}, which aims to enable modern social media and online conversations to function similarly to the early web, where individuals could easily create blogs or use Really Simple Syndication (RSS) feeds to follow multiple blogs. This approach is intended to foster a more open and decentralized online community \cite{Masnick2019Protocols} compared to ``walled-garden'' platforms like Facebook and Twitter/X. 
The AT protocol is an open framework for creating social applications, offering users insight into their construction and development. 
It establishes a standard format for user identity, following mechanisms, and data across social applications, facilitating interoperability between apps and empowering users to transfer their accounts effortlessly.
Thus, Bluesky's model aims to promote competition among developers, who are free to build various interfaces, filters, and additional services.
According to Bluesky, this competitive environment should decrease the necessity for censorship as the best solutions naturally gain prominence~\cite{BlueskyFAQ2024}.

In this paper, we provide the first exploration of how opening the Bluesky platform to the public affected key metrics of interest.
We investigate patterns of user activity, their political leaning, and the quality of information circulating on the platform.
Crucially, we explore these characteristics both before and after the platform was opened to the public.
For a general analysis of the activity on the platform during the first year, we refer the reader to contemporary work \cite{failla2024m,quelle2024bluesky}.

\section{Related work}

Distributed social networks aim for decentralization, allowing users to have more control and privacy. Early efforts like LifeSocial.KOM \cite{Graffi2011LifeSocialKOMAS} and PeerSON \cite{Buchegger2009PeerSoNPS} were based on the peer-to-peer model but faced challenges in performance and reliability.
This led to a shift toward server-based federated models like Mastodon \cite{Raman2019ChallengesIT, Zignani2018FollowT}.
This approach balances flexibility and ease of use while maintaining some decentralization \cite{bono2024exploration}.

Mastodon, created in 2016, is a free and open-source social media platform that allows users to create their own servers (``instances'') and connect with others across the globe.
It is a decentralized social network, meaning that it is not owned by a single entity, but rather a network of independent servers that are connected together. 
A number of studies have identified striking features that make up Mastodon's distinct ``fingerprint,'' distinguishing it from better-known online social networks \cite{LaCava2021UnderstandingTG,bono2024exploration}.
Mastodon has, however, suffered from some natural pressures towards centralization, which can lead to potential points of failure \cite{Raman2019ChallengesIT}.

To overcome this weakness, Bluesky developed its own AT protocol in order to provide decentralization, with several features that distinguish it from other authentication protocols.
Scalability, security, and ease of use make it an attractive option for building open and decentralized social media applications that prioritize user privacy and data security \cite{Kleppmann2024ATProtocol}.
Using standard web technologies and re-using existing data models from the Web 3.0 protocol family also contribute to its efficiency and reliability \cite{edwards2023social}.
Additionally, its federated networking model bolsters security by dispersing data across numerous servers, mitigating the risk of a single point of failure\cite{Raman2019ChallengesIT}.
As mentioned above, two contemporary work on Bluesky have been published: \cite{failla2024m,quelle2024bluesky}. Unlike this work, they provide a general analysis of Bluesky's first-year activities. 

\section{Methods}

The present analysis is based on data collected in a two-month period around the time that the platform opened to the public.

\subsection{Data collection}
\label{data_collection}

We collected data by accessing Bluesky's public and free ``Firehose'' endpoint, which provides developers with real-time access to atomic actions performed on Bluesky such as user posts, follows, likes, etc. \cite{bluesky_firehose,atProtocol_eventStream}.
If a real-time connection with Bluesky is interrupted, the endpoint enables data collectors to resume collection from the moment the connection was lost, retrieving data from up to 72 hours prior.
This ensures comprehensive data coverage throughout the observed timeframe.
We utilized the \texttt{dart} library from the AT protocol \cite{BlueskyDart2024} to fetch data by invoking the \texttt{com.atproto.sync.subscribeRepos} endpoint, also known as the Firehose endpoint \cite{BlueskyFirehose2024}.
This process, facilitated by an existing open-source project \cite{BurghardtFirehose2024}, is straightforward and flexible due to the absence of user authentication requirements in the AT protocol.

\begin{table}[t]
\caption{Basic statistics of the Bluesky dataset.}
\begin{tabular}{lrrrrr}
\hline
&  original  &  reply &  repost &  total & \% \\ 
\hline
messages  & 30,235,716 & 20,842,322 & 20,530,919 & 71,608,957 & \\ 
\hspace{1em} with links & 3,215,963  & 339,921 & 2,683,874  & 6,239,758 & 8.57\%    \\ 
active users & & & &  2,734,569& \\ 
\hspace{1em} sharing messages & & & & 1,752,083  &   \\ 
\hspace{2em} with links & & & & 389,077  & 14.22\%  \\ 
follow actions & & & & 39,214,164 & \\ 
block actions & & & & 2,799,597 & \\ 
\hline
\end{tabular}
\label{table:characteristics}
\end{table}

Bluesky enables tracking of various user activities on the platform.
Our analyses focus on several key actions: posts, replies, reposts, follows, and blocks. 
These actions represent the main forms of user interaction and communication. 
Users follow others to stay updated, create posts to share content or repost content created by others, and reply to engage in discussions. 
While the Bluesky terms of service do not impose privacy restrictions on the data collection, we collect only public information about users, posts, and any attached metadata in accordance with Bluesky's Privacy Policy.\footnote{\url{https://bsky.social/about/support/privacy-policy}} 
We do not make the data publicly available and only provide anonymized information in this paper, except for a few prominent accounts in \S\ref{sec:4.8}.

Table~\ref{table:characteristics} provides basic statistics of our dataset, which spans 56 days, from Jan 9th to March 4th 2024, and comprises 114 million activities  (message actions plus other actions like follow and block).

\subsection{News source labeling}
\label{sec:labeling}

To assess the reliability of news outlets shared on Bluesky, we label web domains using NewsGuard\footnote{\url{https://www.newsguardtech.com/}} ratings, following a consolidated approach in the literature \cite{cheng2021twittervsfacebook,pierri2023one}. 
NewsGuard is a reputable and unbiased organization that employs experts to evaluate news sources based on factors such as transparency, accountability, adherence to journalistic standards, and error correction to determine credibility. NewsGuard ratings range from 0 (highly unreliable) to 100 (highly reliable).
Approximately 6.2 million posts (8.7\% of all posts) contained a URL.
Of these, we were able to label around 1 million (16\% of posts with URLs, and 1.4\% of all posts) with a NewsGuard rating.
The average rating of the posts with links in the social network is high, close to 94. 
We further define low-credibility websites as news outlets with a NewsGuard rating of 30 or lower, following previous literature \cite{Nogara2024Misinfo,pierri2020diffusion,pierri2021vaccinitaly}. 
We also leveraged political bias ratings from Media Bias/Fact Check,\footnote{\url{https://mediabiasfactcheck. com/}} an independent organization that rates news media sources, to label the political leaning of news websites shared on Bluesky. Information sources are categorized across a seven-point political spectrum: Extreme Left (-3), Left (-2), Left-Center (-1), Least Biased (0), Right-Center (1), Right (2), and Extreme Right (3). 

\section{Results}

\subsection{Online activity}

\begin{wrapfigure}[17]{r}{0.6\linewidth}
\vspace{-4em}
\centerline{\includegraphics[width=\linewidth]{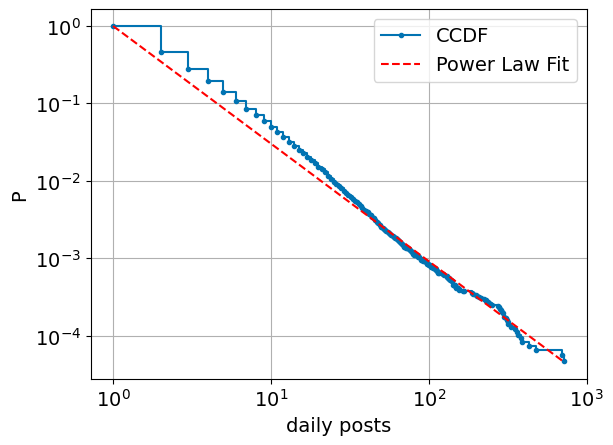}}
\vspace{-1em}
\caption{Complementary cumulative distribution of user activity (number of original posts) on Bluesky during one day (March 4, 2024).
The dashed line is a power-law fit of the  distribution, which yields a slope of 1.53 ($x_{min}=1, x_{max}=1{,}000$).}
\label{fig:power-law}
\end{wrapfigure}

We observe in
Table~\ref{table:characteristics} that original posts are the most common user activity on the platform, indicating that users tend to create more original content rather than reshare or interact with existing posts. This contrasts with centralized social media platforms, such as Twitter/X, where resharing through retweets is more prevalent \cite{alshaabi2021growing,luceri2021down}, and is likely due to the sudden increase in the user-base.

The daily average number of posts per active user is 3.16 (95\% CI: [3.13, 3.20]). 
However, the user activity is quite heterogeneous. Fig.~\ref{fig:power-law} illustrates that the number of daily posts $x$ follows a broad, power-law distribution $P(x) \sim x^{-\alpha}$ with $\alpha \approx 2.53$. This exponent value is consistent across different days, ranging between 2.50 and 2.71.
This extreme level of heterogeneity indicates that the average is not a good statistical descriptor of the activity distribution, as the fluctuations are very high: most users post infrequently but a non-negligible fraction posts several hundred messages per day.

\subsection{Temporal patterns}

\begin{figure}[b!]
    \centering
    \begin{floatrow}
      \ffigbox[\FBwidth]{\caption{User activity on Bluesky before and after the opening. The vertical dashed  line represents the date of the opening (Feb 6).}\label{fig:before-after-users}}{%
        \includegraphics[width=\linewidth]{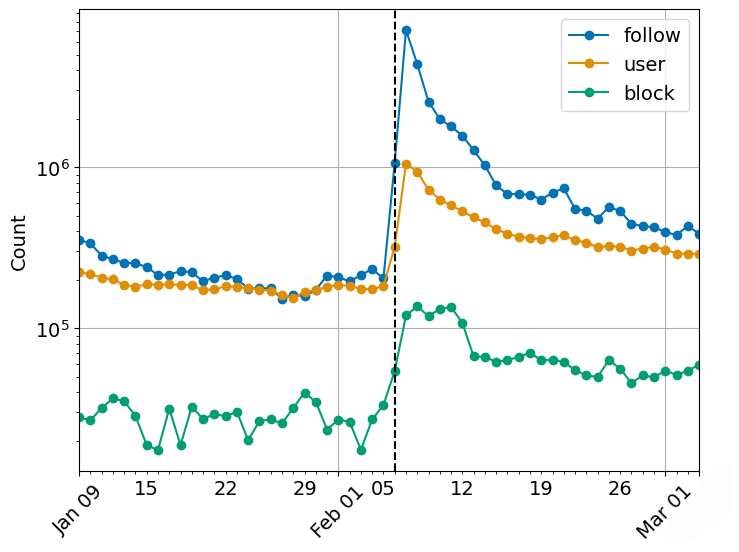} 
      }
      \ffigbox[\FBwidth]{\caption{Sharing activity on Bluesky before and after the opening. The vertical dashed line represents the date of the opening.}\label{fig:before-after-posts}}{%
        \includegraphics[width=1.02\linewidth]{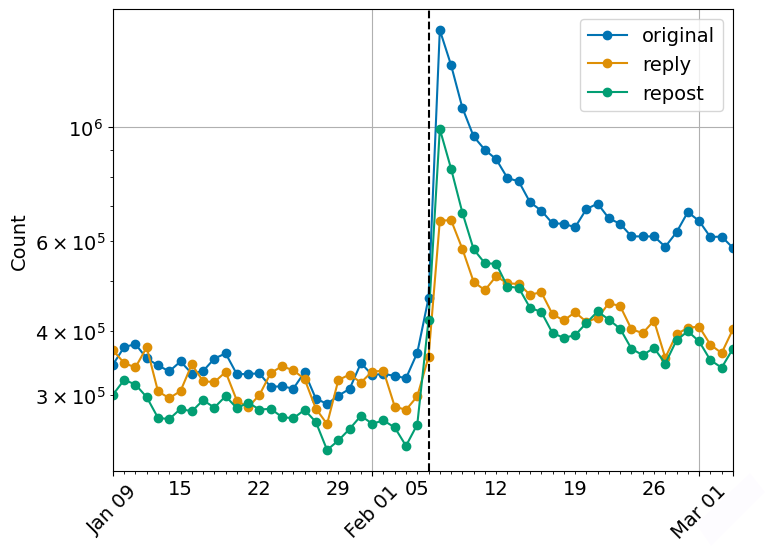}
      }
    \end{floatrow}
\end{figure}

To investigate the impact of the Bluesky opening on its user base, Fig.~\ref{fig:before-after-users} plots the daily number of active users and following actions. The platform's opening on February 6th (dashed line) resulted in spikes of activity, up to 1 million active users and over 7 million follow actions on the following day.
These spikes in behavior represent an almost six-fold increase in active users and an almost 35-fold increase in following actions, relative to the day before the opening.
Both trends decreased rapidly in the following days, stabilizing at levels slightly higher than those seen before the opening, likely as the initial excitement around the new platform subsided.

During the entire observation period, 287,539 users blocked a total of 758,681 accounts, yielding a ratio of 2.7 block actions per user. The average number of block actions per user is higher (9.5) because this blocking activity is heterogenous, with some users blocking many accounts.  
Fig.~\ref{fig:before-after-users} shows that an increase in users and following activities is associated with a rise in blocking activities, with a 3.6-fold increase from 33,269 to 120,054 instances of blocking. 
After a few days, the frequency of blocking activities stabilized but remained slightly higher than the levels observed before February 6.

Fig.~\ref{fig:before-after-posts} illustrates the volume of shared posts over time.
We distinguish user-sharing activities into original posts, replies, and reposts, as described in \S\ref{data_collection}. 
Table~\ref{table:characteristics} previously highlighted that original posts are the most prevalent type of shared activity, a trend that becomes particularly prominent following the platform's public launch.

As expected, we observe a significant increase in the volume of shared content coinciding with the platform's opening on February 6th.
This surge is reflected in all types of sharing activities, likely due to the influx of new users joining during this period, as shown in Fig.~\ref{fig:before-after-users}. 
Specifically, the volume of original posts increased approximately 4.3 times, from 362k on February 5th to 1.5M the following day.
Reposts and replies also rose from 262,201 and 297,717 to 990,835 and 656,063, respectively.
Despite these substantial initial increases, the volume of each activity type diminished in the subsequent days. 

\subsection{Languages}
\label{language_analysis}

\begin{wrapfigure}[13]{r}{0.65\linewidth}
\vspace{-4.5em}
\centerline{\includegraphics[width=\linewidth]{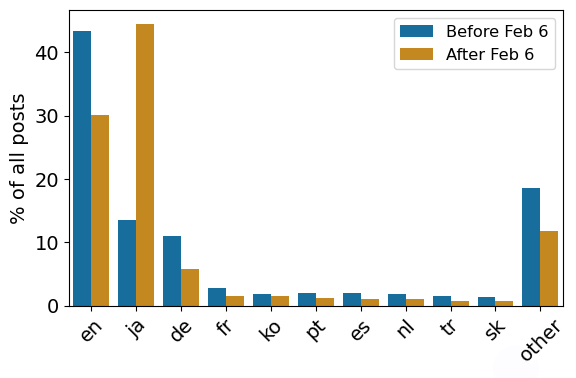}}
\vspace{-1em}
\caption{Top 10 languages in Bluesky. Bars of the same color sum to one.}
\label{fig:language}
\end{wrapfigure}

To examine the language distribution in user-shared content, we removed irrelevant text (e.g., URLs, emojis, etc.) and applied a language classifier using the \texttt{langdetect}\footnote{\url{https://pypi.org/project/langdetect/}} NLP library.

Fig.~\ref{fig:language} displays the prevalence of the top 10 languages in user posts, highlighting a dominance of English and Japanese, which together comprise more than two-thirds of all content. 
Prevalence of English content decreased from 43\% to 30\% after the platform's opening, while
the share of Japanese content increased from 14\% before the opening to 44\% afterward. 

\begin{wrapfigure}[15]{r}{0.65\linewidth}
\vspace{-3em}
\centerline{\includegraphics[width=\linewidth]{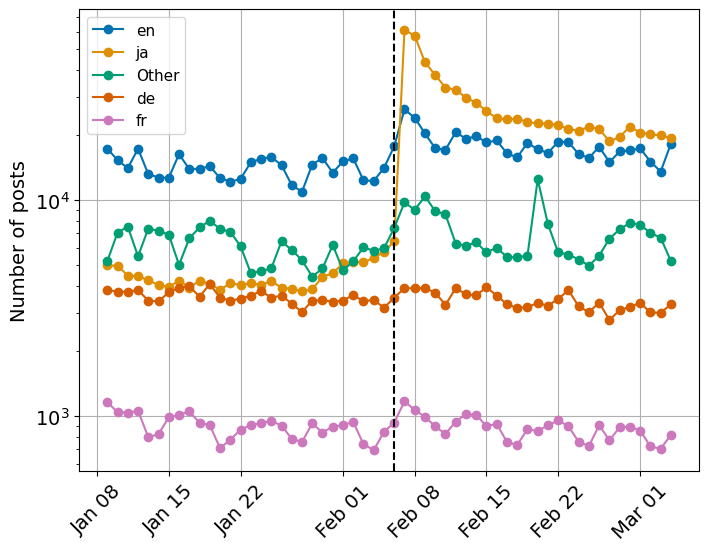}}
\vspace{-1em}
\caption{Trend of 5 top languages on Bluesky during the observation period.}
\label{fig:language-trend}
\end{wrapfigure}

Fig.~\ref{fig:language-trend} illustrates the trends of the top five languages used during the observation period.
We note a significant increase in Japanese posts just after the opening, making it the most used language.
Meanwhile, content in other languages, including English, remains relatively similar to how it was before the opening, experiencing only small fluctuations following the opening.

\subsection{Information sources}
\label{sec:Information Sources Changes}

\begin{figure}[t]
         \centering
         \includegraphics[width=.49\columnwidth]{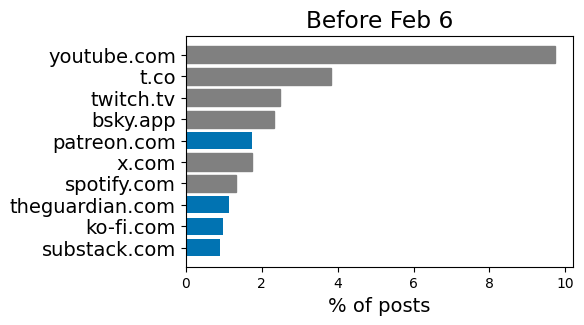}
         \includegraphics[width=.49\columnwidth]{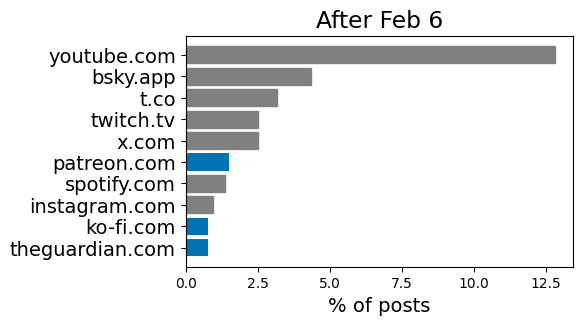}
\caption{Websites shared on Bluesky: in blue, we highlight the websites that are neither social media nor shortened links.}
\label{fig:urls-wSM}
\end{figure}

Fig.\ref{fig:urls-wSM} illustrates the ten domains most shared before and after February 6.
We observe that the most common links are either to other social media platforms, Bluesky itself, or link-shortening services like \url{bit.ly}  and \url{t.co}, Twitter's built-in service.
This appears to suggest that a great deal of content on Bluesky is in reference to other, more mainstream platforms.

\begin{figure}[t]
         \centering
         \includegraphics[width=.48\columnwidth]{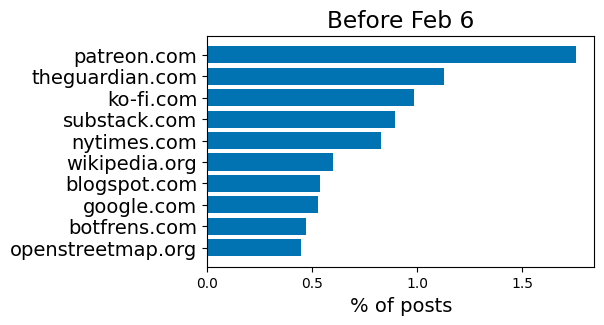}
         \includegraphics[width=.48\columnwidth]{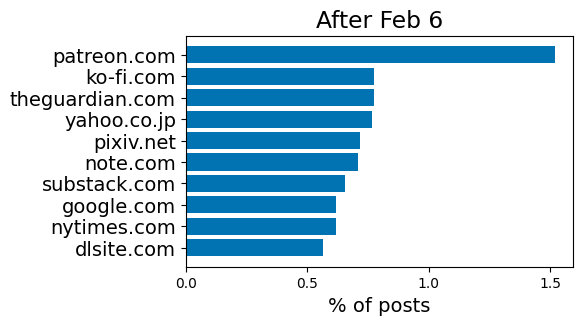}
\caption{Websites shared on Bluesky excluding social media and shortened links.}
\label{fig:urls-noSM}
\end{figure}

Fig.~\ref{fig:urls-noSM} presents the same information after removing links to social media platforms (spotify.com, twitter.com, x.com, twitch.tv, youtu.be, youtube.com, bsky.app, instagram.com) and link shorteners (t.co, bit.ly, t.me).
This procedure allows us to observe that content from news and information outlets like \textit{The Guardian}, \textit{The New York Times}, and \textit{Wikipedia} constituted a larger share of the content before the platform's opening to the public.
Consistent with the language analysis (\S\ref{language_analysis}), we observe a surge in Japan-related activity, with a high percentage of links to Japanese websites following the opening (yahoo.co.jp, pixiv.net, note.com, and dlsite.com). 
Patreon and Ko-fi, two platforms for crowdfunding and content creation, rank as the most shared web domains beyond social media and news platforms. While such sites have been used to launch fundraising campaigns in times of distress \cite{ye2023online}, manual inspection reveals that Patreon is primarily used to promote adult content on Bluesky.

\newpage
\subsection{Political leaning}

\begin{wrapfigure}[18]{r}{0.65\linewidth}
\vspace{-4em}
\centerline{\includegraphics[width=\linewidth]{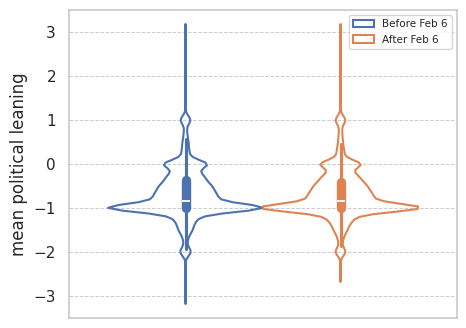}}
\vspace{-1em}
\caption{Political leaning of Bluesky users (with at least 5 rated posts) before and after February 6. Leanings are calculated by averaging the political alignment of websites shared by each user.  
The violin plots also display the interquartile range (box) and median (horizontal line inside the box).}
\label{fig:political}
\end{wrapfigure}

Fig.~\ref{fig:political} shows the distribution of the estimated political alignment of active Bluesky users, defined as those engaging in at least five posts with links to rated websites.
The political alignment of each user is obtained by averaging the political alignment of the websites they share during the observation period (see \S\ref{sec:labeling}). 
We used a Mann-Whitney test and did not find a statistically significant difference between the distributions of the political leanings before and after the opening ($p=0.05$). 

\begin{wrapfigure}[18]{r}{0.82\linewidth}
\vspace{-2em}
\centerline{\includegraphics[width=\linewidth]{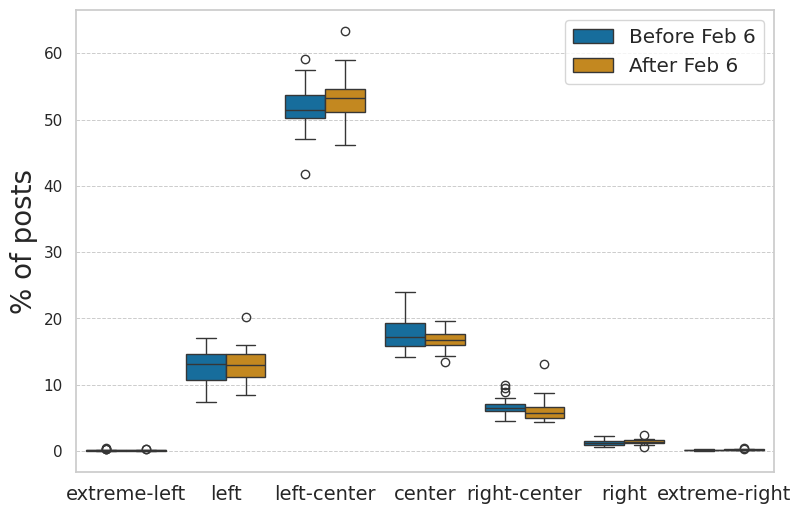}}
\caption{Political leaning of Bluesky posts before and after Feb. 6.}
\label{fig:political-posts}
\end{wrapfigure}

Similarly, we analyzed the distribution of the estimated political alignment of posts (Fig.~\ref{fig:political-posts}). 
While no significant difference was found at the user level before and after the opening, some variation is observed at the post level, particularly among right-center (Mann-Whitney: $p = 0.03$), right ($p = 0.046$), and extreme-right posts ($p = 0.008$).

\subsection{Credibility}

The number of users sharing any link with a credibility rating is 36,713 before and 47,612 after the opening. 
A small fraction of these (0.12\%) shared low-credibility content. 

\begin{figure}
\centerline{\includegraphics[width=.7\linewidth]{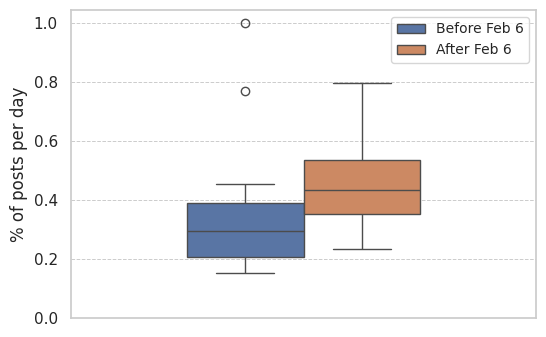}}
\caption{Distributions of the percentage of posts linking to low-credibility sources, out of all posts linking to rated sites, for each day in the periods before and after the platform opened.}
\label{fig:credibility-score-NG}
\end{figure}

Fig.~\ref{fig:credibility-score-NG} illustrates the production of low-credibility content before and after the opening.
While the volume of this content on Bluesky remained low (around 0.4\% of all posts on an average day), we find that it significantly increased after the opening (Mann-Whitney: $p<0.001$). 

\begin{figure}
\centerline{\includegraphics[width=0.8\linewidth]{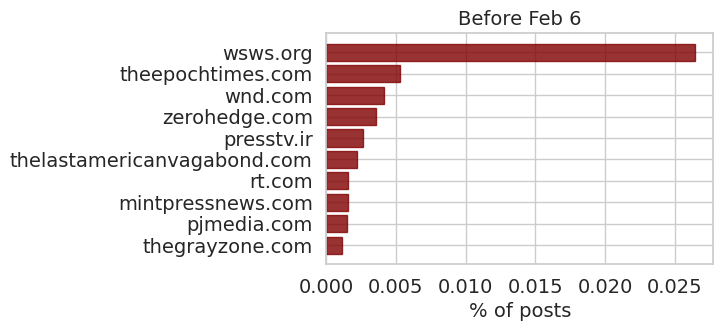}}
\caption{Most shared low-credibility websites before Feb. 6.}
\label{fig:credibility-websites}
\end{figure}

\begin{figure}
\centerline{
        \includegraphics[width=0.4\columnwidth]{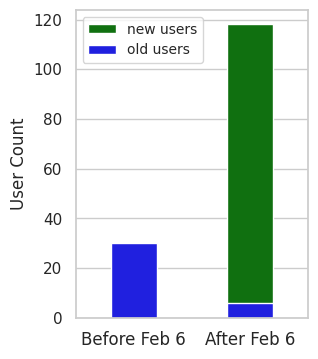}
        \includegraphics[width=0.4\columnwidth]{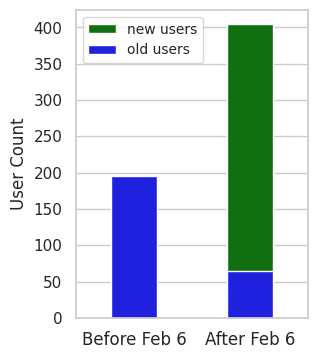}
}
\caption{Number of users sharing links with credibility ratings before and after the opening for different subsets of users. Left: Users who only shared links to low-credibility sources. Right: Users who shared at least one low-credibility link.}
\label{fig:credibility-users}
\end{figure}  

Fig.~\ref{fig:credibility-websites} reports on the most shared low-credibility websites before the opening. 
The list remained similar after the opening, with the notable exception of the top website, \url{wsws.org}, whose shares decreased to half of the percentage recorded prior to the opening.
This is not due to a decrease in the share of this domain, but rather to the increase in the share of all domains.

Similar to what has been reported for Twitter \cite{DeVerna2022FIB,Nogara2022TheDD,pierri2021vaccinitaly,Grinberg2019Jan}, we identified a small subset of users responsible for most of the unreliable content shared in Bluesky's brief existence. 
In particular, we observe that ten accounts (1.8\% of all users who spread low-credibility content) are responsible for spreading 62\% of links to low-credibility sources.
Manual inspection of each account confirms that all were created before the platform opened to the public and remain active at the time of writing.
Furthermore, they exhibit suspicious behavior, posting a high volume of content almost exclusively from low-credibility news outlets, possibly in automated fashion. 

F~\ref{fig:credibility-users} shows the number of users sharing links to low-credibility sources before and after the opening, distinguishing between new and existing users. 
We observe that the number of users who only shared low-credibility links quadrupled after the opening.
The number of those who shared at least one low-credibility link approximately doubled.

\subsection{Toxicity}

We used the Perspective API \cite{lees2022new} to analyze the toxicity of shared content in both English and Japanese. 

\begin{figure}[t]
\centerline{\includegraphics[width=0.8\linewidth]{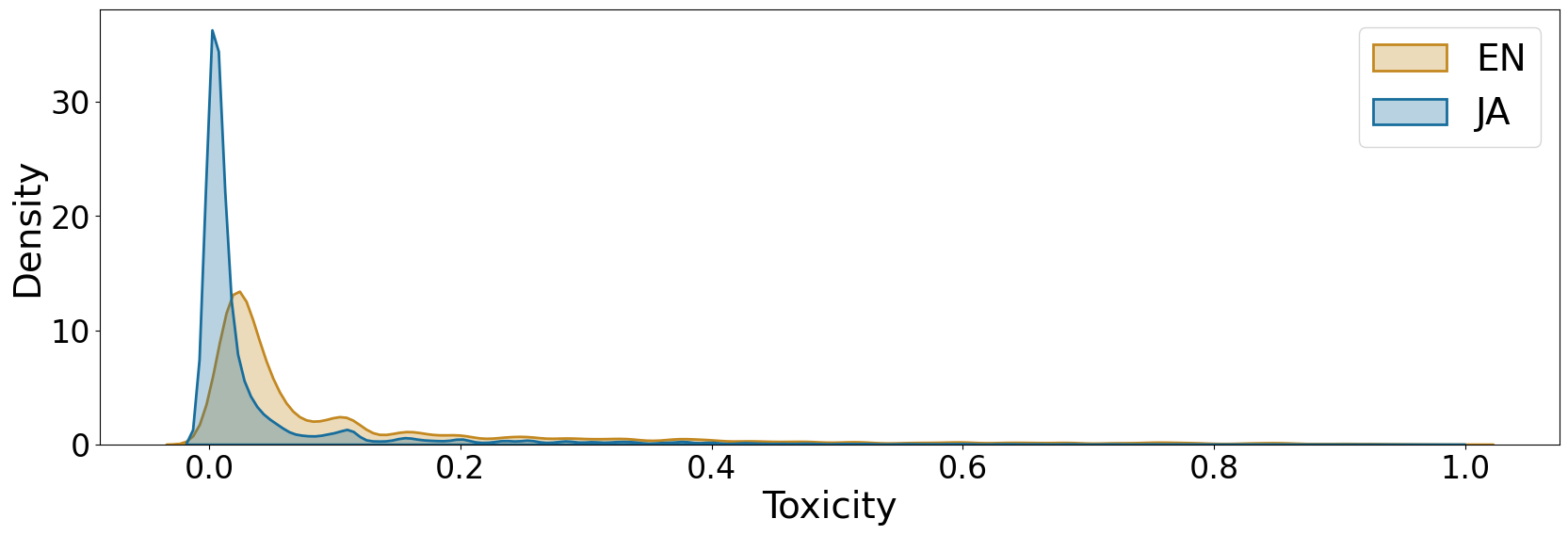}}
\centerline{\includegraphics[width=0.8\linewidth]{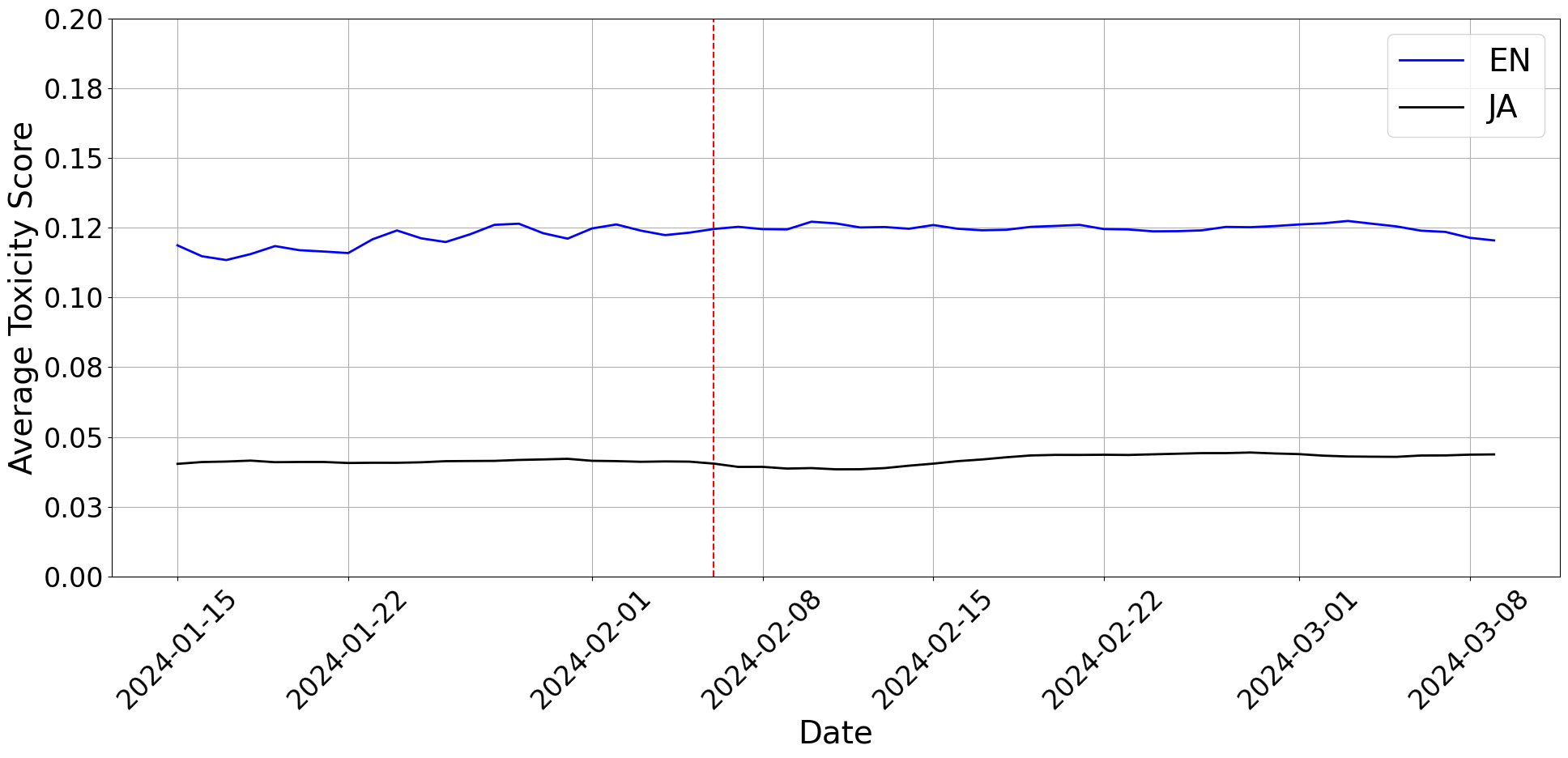}}
\caption{Toxicity of Bluesky posts in English (EN) and Japanese (JA). Top: Distribution of toxicity for the entire observation period. Bottom: Weekly running average of daily toxicity scores. The dashed red line indicates the platform opening.}
\label{fig:toxicity}
\end{figure}

As illustrated in Fig.~\ref{fig:toxicity} (top), English content tends to have higher toxicity scores.
Fig.~\ref{fig:toxicity} (bottom) shows that while, on average, the toxicity of English content is three times higher than that of Japanese content, the values remain stable over time.

\subsection{Follower network}
\label{sec:4.8}

\begin{figure}[b!]
    \centering
    \begin{floatrow}
      \ffigbox[\FBwidth]{\caption{Complementary cumulative out-degree distributions of node in the follower network.}\label{fig:out-degree}}{%
        \includegraphics[width=\linewidth]{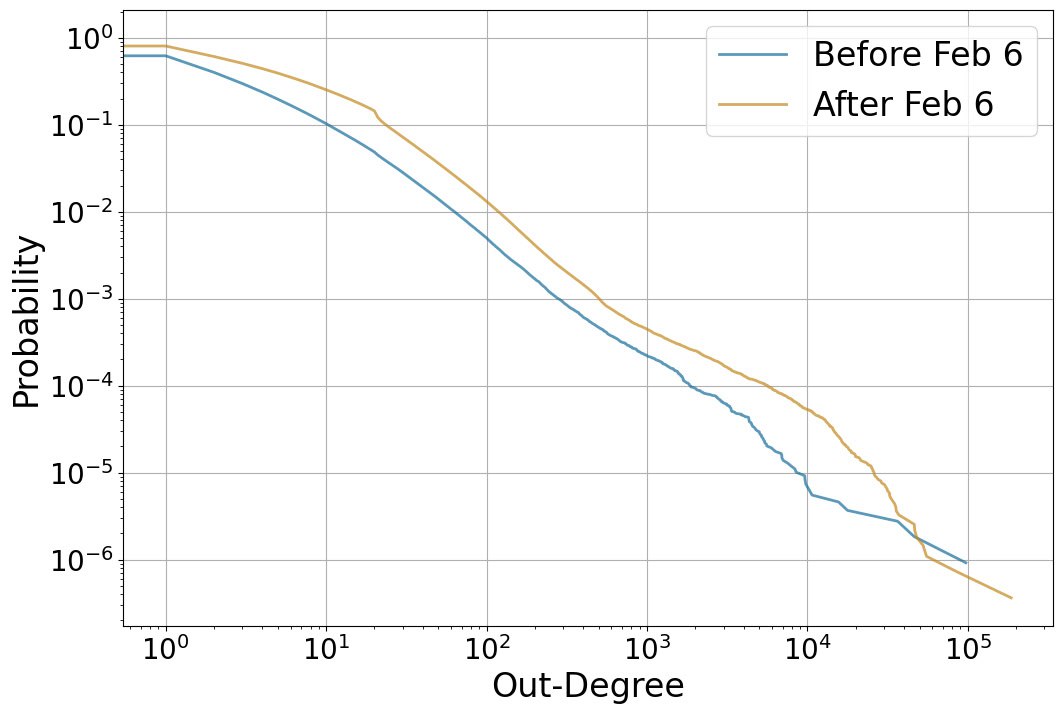} 
      }
      \ffigbox[\FBwidth]{\caption{Complementary cumulative in-degree distributions of node in the follower network.}\label{fig:in-degree}}{%
        \includegraphics[width=\linewidth]{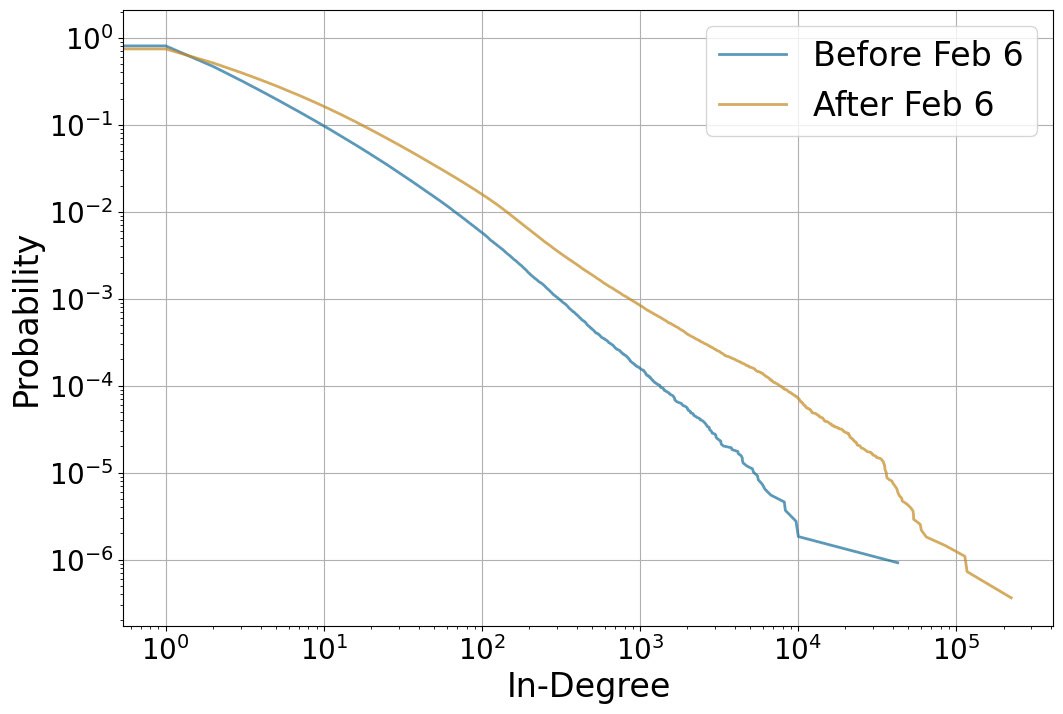}
      }
    \end{floatrow}
  \end{figure}

In Fig.~\ref{fig:before-after-users}, we observed a significant increase in the \textit{following} activities.
This increase is reflected in various follower network statistics before and after the opening, as detailed in Table~\ref{table:network-statistics}.
While the density of the follower network slightly decreased after the opening, the size of the strongly connected component more than tripled, and the average degree more than doubled, indicating that Bluesky users tend to follow more accounts after the opening. The out-degree distributions in Fig.~\ref{fig:out-degree} confirm this trend.
\setlength{\tabcolsep}{12pt}
\begin{table}[t]
\caption{Follower network statistics. LSCC stands for the largest strongly connected component.}
\centering
\label{table:followers}
\begin{tabular}{lccc}
\cline{2-4}
                 & Before Feb. 6 & After Feb. 6  & Difference      \\ \hline
Number 			of nodes & 1,088,539 & 2,751,272  &   1,662,733       \\
Number 			of edges & 5,230,054 & 28,838,739  & $-$23,608,685         \\  
Density & $4.4 \times 10^{-6}$    &  $3.8 \times 10^{-6}$ & $- .6 \times 10^{-6}$\\
Avg. in-degree       & 9.6     & 20.3 &   10.7           \\
Avg. out-degree       & 7.4     & 14.9 &   7.5           \\
LSCC size      & $\sim$200k  & $\sim$650k &     $\sim$450k       \\
\hline
\end{tabular}
\label{table:network-statistics}
\end{table}
\setlength{\tabcolsep}{6pt}

\begin{table}[t]
\caption{Annotated list of the ten users with the most new followers and the ten users who performed most follow activities before and after the platform's opening. Only the real names of prominent accounts are included for privacy reasons.}
\centerline{
\begin{tabular}{lr|lr||lr|lr}
\multicolumn{4}{c||}{Accounts with most new followers} &\multicolumn{4}{c}{Accounts who followed most accounts}     \\ \hline 
\multicolumn{2}{c|}{Before Feb. 6} & \multicolumn{2}{c||}{After Feb. 6} &  \multicolumn{2}{c|}{Before Feb. 6} & \multicolumn{2}{c}{After Feb. 6} \\ \hline 
Account    & Num.    & Account    & Num.   &   Account    & Num.    & Account    & Num.        \\ \hline
Bluesky \scriptsize{\faArrowsAltH}    & 43,801& Bluesky \scriptsize{\faArrowsAltH}  & 120,709  & user & 50,570             & user \scriptsize{\faCalendar[regular]~\faRadiation} & 167,220   \\   
user \scriptsize{\faArrowsAltH} & 10,256 & user \scriptsize{\faCalendar[regular]~\worldflag[width=5pt]{JP}} &88,438&user&24,588& user \scriptsize{\faCalendar[regular]~\worldflag[width=5pt]{JP}}&85,075\\
user &    9,886      & user \scriptsize{\faArrowsAltH}&66,535&user \scriptsize{\faUserTimes}&23,482&user \scriptsize{\faCalendar[regular]~\worldflag[width=5pt]{JP}} &55,073\\
Wash. Post \scriptsize{\faArrowsAltH}&8,508&NY Times \scriptsize{\faArrowsAltH}&62,087&user &21,907& user \scriptsize{\faCalendar[regular]~\worldflag[width=5pt]{JP}} &45,848\\
NY Times \scriptsize{\faArrowsAltH}&8,393&Wash. Post \scriptsize{\faArrowsAltH}&60,904&user &13,414&user \scriptsize{\faCalendar[regular]~\faUserMinus} &44,588\\
user&6,809&user&55,081&user&12,310&user \scriptsize{\faUserTimes}&44,460\\
Bluesky CEO&6,314&user \scriptsize{\faCalendar[regular]~\worldflag[width=5pt]{JP}}&54,944&user&10,689&user \scriptsize{\faCalendar[regular]~\faUserMinus} &43,786\\
user \scriptsize{\faArrowsAltH~\faNewspaper[regular]}&5,744&Bloomberg \scriptsize{\faArrowsAltH}&54,800&user&9,950&user \scriptsize{\faCalendar[regular]~\faRadiation}&40,926\\
user \scriptsize{\faArrowsAltH~\faNewspaper[regular]}&5,621&user \scriptsize{\faArrowsAltH~\faNewspaper[regular]}&52,455&user&8,211& user \scriptsize{\faCalendar[regular]~\worldflag[width=5pt]{JP}}&37,209\\
Bloomberg \scriptsize{\faArrowsAltH}&5,332&user \scriptsize{\faCalendar[regular]~\worldflag[width=5pt]{JP}}&49,697& user&7,588& user \scriptsize{\faCalendar[regular]~\worldflag[width=5pt]{JP}}&36,189\\ \hline
\multicolumn{8}{l}{Account type: \textsuperscript{\faRadiation}spam; \textsuperscript{\faUserMinus}deleted; \textsuperscript{\faUserTimes}suspended; \textsuperscript{\worldflag[width=5pt]{JP}}Japanese; \textsuperscript{\faNewspaper[regular]}Journalist.}\\
\multicolumn{8}{l}{Time key: \textsuperscript{\faCalendar[regular]}created after Feb. 6; \textsuperscript{\faArrowsAltH}present before and after Feb. 6.}
\end{tabular}
}
\label{table:follower-followee}
\end{table}
Table~\ref{table:follower-followee} presents the ten accounts that gained the most new followers alongside the ten accounts that followed the most other users before and after the platform's opening.
For comparison, the average number of followers before and after February 6th is 9.6 and 20.3, respectively, while the median is 3.0 and 4.0, respectively. 

The average number of accounts followed by Bluesky users before and after February 6th is 7.4   and 14.9, respectively, while the median is 2.0 and 4.0, respectively.
This analysis points to a potentially spammy behavior: some accounts followed a suspiciously large number of users right after joining the Bluesky, as illustrated in Fig.~\ref{fig:out-degree}.

The crossover in the tail (just before degree $10^{5}$) in the out-degree CCDF curves is due to the uneven distribution of follow activities and to the discrepancy in the number of users before and after the opening. In fact, both before and after the opening, we have only one outlier user who performed a high number of follow actions  (respectively, 50,570 and 167,220 actions, which is about 2-3 times the number of actions of other users in Table~\ref{table:follower-followee}). Furthermore, since the users who performed follow actions before the opening are about $1/3$  compared to the users who performed follow actions after (see Table~\ref{table:network-statistics}), the probability in the CCDF tail for the first 
is higher than the one for the latter. When we remove these two outlier users, this crossover between the two CCDF curves disappears.

The CCDF curves for the in-degree distributions before and after the opening are illustrated in Fig. \ref{fig:in-degree}.  Table~\ref{table:follower-followee} report the ten top users for both periods.

Among the users that gained the most followers, several are consistent across both periods, with the majority being news outlets, such as \textit{Washington Post}, \textit{New York Times}, and \textit{Bloomberg}.
This prevalence of news outlets among the accounts with the most new followers suggests that Bluesky may be evolving into another platform for news dissemination, potentially replacing Twitter.
However, during the observed period, three new Japanese accounts emerged in this category, with one reaching the second position, behind the official Bluesky account. 

Before the platform's opening, the users who followed the most new accounts were primarily English-language accounts and appeared to engage in normal activity, except for one user suspended by Bluesky.
However, after the opening, the composition of these users changed significantly: half were Japanese, two accounts were deleted, one was suspended, and two were classified as spam by Bluesky. 
Unlike the overall network behavior, these users were more engaged in reposting activities than in creating original posts.

\section{Discussion}

We provided the first large-scale analysis of how opening the Bluesky platform to the public affected user activity and network structure.
We observe a broad distribution of activity similar to that of more established platforms; however, there is a higher volume of original content compared to reshared content.
This, coupled with the very low toxicity we observed, contrasts with other (centralized) social media platforms. 
After opening to the public, Bluesky experienced a large surge in new users, especially posting English and Japanese content, and activities.
The significant increase in Japanese users deserves further investigation; it might signify a  conversational system that is more appealing to this population. 

We observed no significant changes in the toxicity level and in the political leaning of users, but some variation is observed at the post level.
Further, we discovered some users exhibiting suspicious behavior, such as connecting with many other users and sharing content from low-credibility news outlets.
We also identified a small subset of users responsible for the majority of unreliable content shared, echoing similar findings on platforms like Twitter/X. 
Additionally, after the platform's opening, a subset of users that followed the most new accounts during this time were flagged as spam or were even deleted or suspended.
These findings suggest some attempts to misuse Bluesky.
However, the fact that some of the suspicious and misbehaving actors have already been banned or tagged as spam indicates that content moderation is taking place on Bluesky.

This work has some limitations. 
First, it is based on a relatively short period (56 days), during which a significant change occurred—allowing anyone to join. 
Second, it was not possible to trace 8\% of the reposts back to their original posts as they predated our data collection period. 
Third, the websites for which we have credibility and political bias scores make up a small fraction of all links shared.  
Finally, our analysis of toxicity is limited only to English and Japanese content. 

Future work could incorporate a more extended multi-language analysis through tools like Perspective API, which can evaluate toxicity in various languages \cite{nogara2023toxic}, as well as longitudinal analyses of user behavior and network structure. 

\section*{Acknowledgments}

This work was partially supported by the Swiss National Science Foundation (grant number CRSII5\_209250) and the Italian Ministry of Education (PRIN PNRR grant CODE prot. P2022AKRZ9 and PRIN grant DEMON \linebreak prot. 2022BAXSPY).

\bibliographystyle{splncs04}
\bibliography{references}

\end{document}